\documentstyle[11pt,epsf,twoside]{article}
\newcommand{\lesssim}{\mbox{\raisebox{0.45ex}{$<$}\hspace{-11pt}
\raisebox{-0.45ex}{$\scriptstyle{\sim}$}}}

\input epsf   

\begin{document}
\setcounter{page}{135}
\title{ Measurement of the Higgs self-coupling at JLC
} 
\author{Y. Yasui$^{1\dagger}$\thanks{e-mail address:yoshiaki.yasui@kek.jp;
This work  was supported in part by Grand-in-Aid for Scientific
Research  $^\dagger$(A)(1) (No. 13047101), $^\ddagger$(C)
(No. 13640309),
$^\S$(A)(No.14046205).
}, 
S. Kanemura$^{1\dagger}$,
S. Kiyoura$^{1,2\dagger}$, 
K. Odagiri$^{1\dagger}$, 
Y. Okada$^{1\ddagger}$, 
E. Senaha$^{1,3}$\\
and S. Yamashita$^{4\S}$
\\
\\
        $^1$
{\it KEK, Tsukuba,1-1 Oho, Tsukuba, Ibaraki 305-0801,Japan}
\\
$^2$
{\it Department of Radiological Sciences, Ibaraki Prefectural
University}
\\
{\it of Health Sciences, Ami, Inashiki, Ibaraki 300-0394, Japan}
\\
        $^3$
{\it Department of Particle and Nuclear Physics, the Graduate University}
\\{\it for Advanced Studies, Tsukuba, Ibaraki 305-0801,Japan
}
\\
        $^4$ {\it 
ICEP, 
University of Tokyo, 7-3-1 Hongo, Bunkyo-ku, Tokyo 113-0033, JAPAN }
}
\date{}
\maketitle

\vspace{-1cm}
\begin{abstract}

\vspace{-0.3cm}
We examine the double Higgs production process at JLC. 
We focus our attention on the measurement of the Higgs self-coupling. 
The sensitivity of the triple Higgs coupling measurement is discussed
in the Higgs mass range 100-200 GeV and the center of mass energy 
to be 500 GeV-1.5 TeV. 
\end{abstract}

\vspace{-0.5cm}
After the discovery of the Higgs boson, detailed study of the Higgs 
potential will be one of the important issues at the linear collider
(LC) experiment\cite{BOU,DJO}.  
From the standard model (SM) Higgs potential 
$V(\Phi)=-\mu^2 \Phi^\dagger \Phi + \lambda(\Phi^\dagger\Phi)^2$,    
we obtain the simple relation between the Higgs boson mass 
$m_h$ and the self coupling $\lambda$ as $m_h^2=2\lambda v^2$,  
where $v$ is the vacuum expectation value of the Higgs boson.  
The precision test of this relation could 
give clear evidence for the new physics\cite{KANE}. 
At the LC, multiple Higgs boson production processes will 
provide the direct information of the triple Higgs coupling.  
In this talk, we discuss the double Higgs production at
JLC\cite{JLC}.   
We examine the potential of the triple Higgs coupling measurement. 
 
At JLC, we can use the two modes, $e^+e^-\rightarrow hhZ$ and 
$e^+e^-\rightarrow (WW)\nu\bar{\nu}\rightarrow hh\nu\bar{\nu}$.  
In the low center of mass energy region($\sqrt{s}~\lesssim~1$ TeV), 
the $hhZ$ mode is dominant.  
On the other hand, the $hh\nu\bar{\nu}$ mode will be important 
at the 1 TeV or higher energies. 
Here, we introduce the effective coupling which includes 
the anomalous contribution as $V(h^3)=(\lambda_3+\delta\lambda_3)hhh$,  
where $\lambda_3(\equiv \lambda v)$ is the SM triple Higgs coupling and 
$\delta\lambda_3$ is the deviation from the SM prediction.  
In figure 1, we show the $\delta\lambda_3$ dependence of the total cross 
section for (a) $e^+e^-\rightarrow hhZ$ and 
(b) $e^+e^-\rightarrow hh\nu\bar{\nu}$.  
Here, the Higgs mass is 120 GeV with the CM energy of 500 GeV, 
1 TeV and 1.5 TeV.  
To compute the cross section, we mainly employ the GRACE\cite{GRC}
system, and use the CompHEP\cite{COMH} system for the cross-check. 
In the low values of the invariant mass $M_{hh}$ for $hh$, 
the $hhZ$ mode is sensitive to $\delta\lambda_3$(Fig. 2). 
We found that the cut for $M_{hh}$ is powerful 
to improve the sensitivity. 
The polarized electron beam is also useful to 
reduce the back-ground for the $hh\nu\bar{\nu}$ mode, 
due to the $V-A$ nature of the weak bosons. 

We show the sensitivity of the triple Higgs coupling 
to $\delta\lambda_3$ in the mass range $100-200$ GeV 
at the parton level (Fig. 3).  
Here, we assumed that the efficiency of the particle tagging is 100\% 
with an integrated luminosity of 1 $ab^{-1}$.  
Dashed (dotted, sold) lines are the results of the $hhZ$ mode
($hh\nu\bar{\nu}$ mode, combined both modes). 
At $\sqrt{s}=500$ GeV, the $hhZ$ mode is dominant. 
If the Higgs mass is relatively light ($m_h~\lesssim$ 150 GeV), 
we can measure the triple Higgs coupling in about 20\% accuracy.  
For $\sqrt{s}$ to be 1 TeV or higher, 
the $hh\nu\bar{\nu}$ mode is dominant.  
Here, we used the invariant mass cut as $M_{hh}<$ 600 GeV for $hhZ$ mode.   
We also used the 100\% polarized electron beam for the $hh\nu\bar{\nu}$
mode. We can expect the high sensitivity in this case
($\delta\lambda_3/\lambda_3~\lesssim~10\%$).  

\vspace{-0.3cm}
{\bf Acknowledgment}
We would like to thank Y. Kurihara, T. Kaneko,
T. Ishikawa, J. Fujimoto, Y. Shimizu 
and S. Kim for valuable discussions.

\begin{figure}[h]
\raisebox{5cm}{(a)}
\epsfxsize=5cm
\epsfysize=4cm
\epsffile{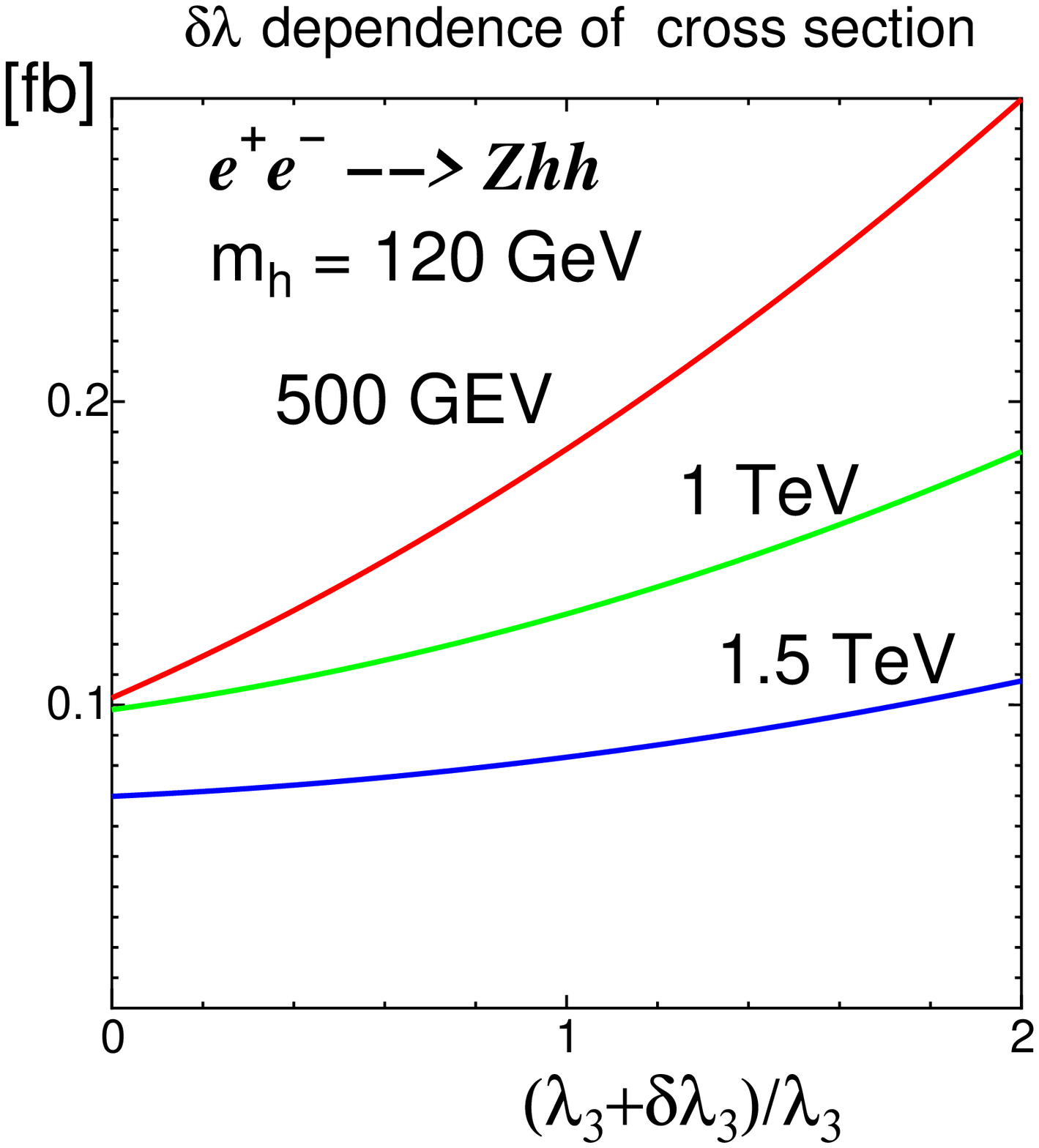}
~~~~~
\raisebox{5cm}{(b)}
\epsfxsize=5cm
\epsfysize=4cm
\epsffile{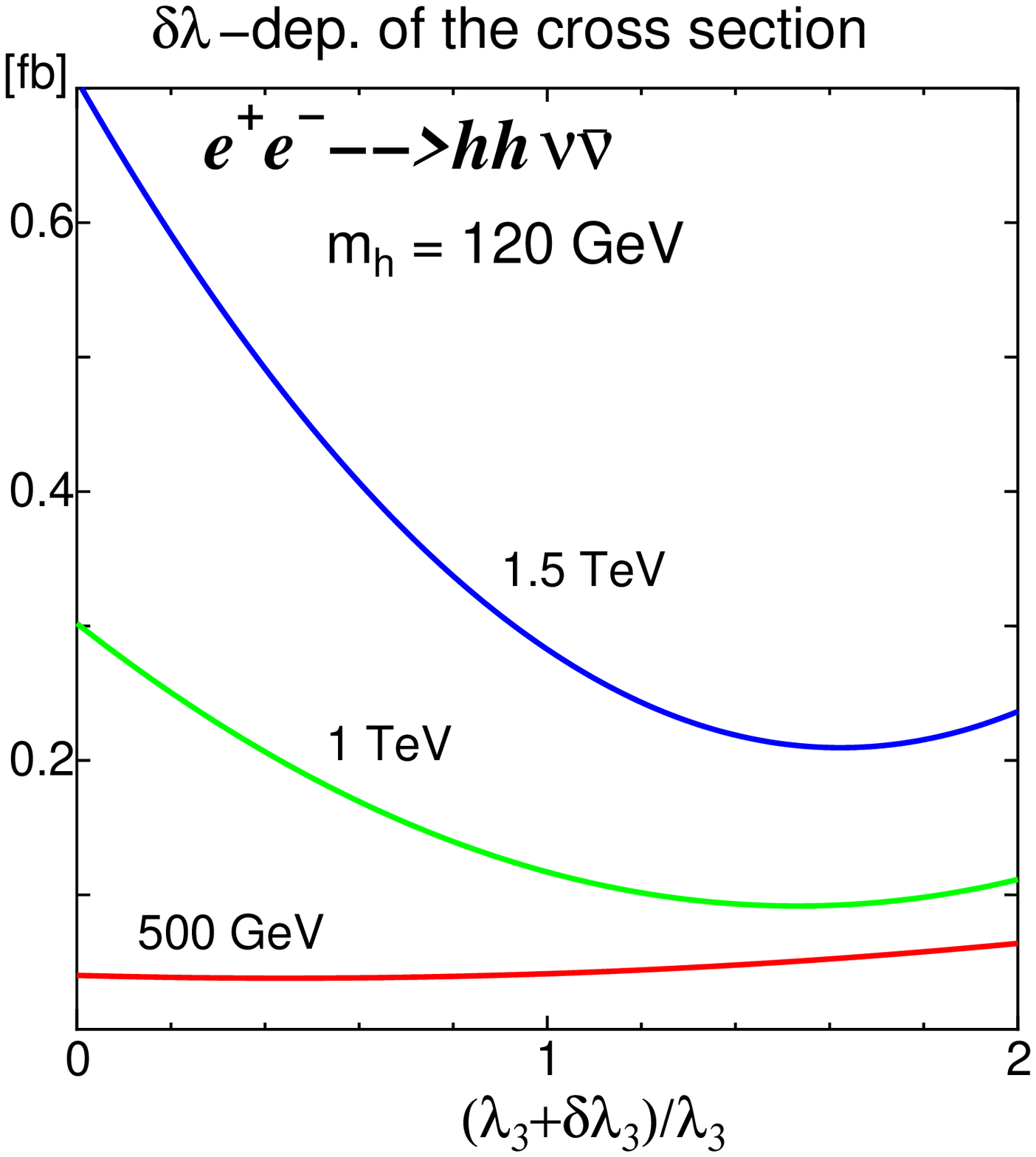} 
\caption{The $\delta\lambda_3$ dependence of the cross section;
(a) $e^+e^-\rightarrow hhZ$, (b)$e^+e^-\rightarrow hh\nu\bar{\nu}$
 }
\end{figure}

\begin{minipage}{.4\linewidth}
\epsfxsize=4.5cm
\epsffile{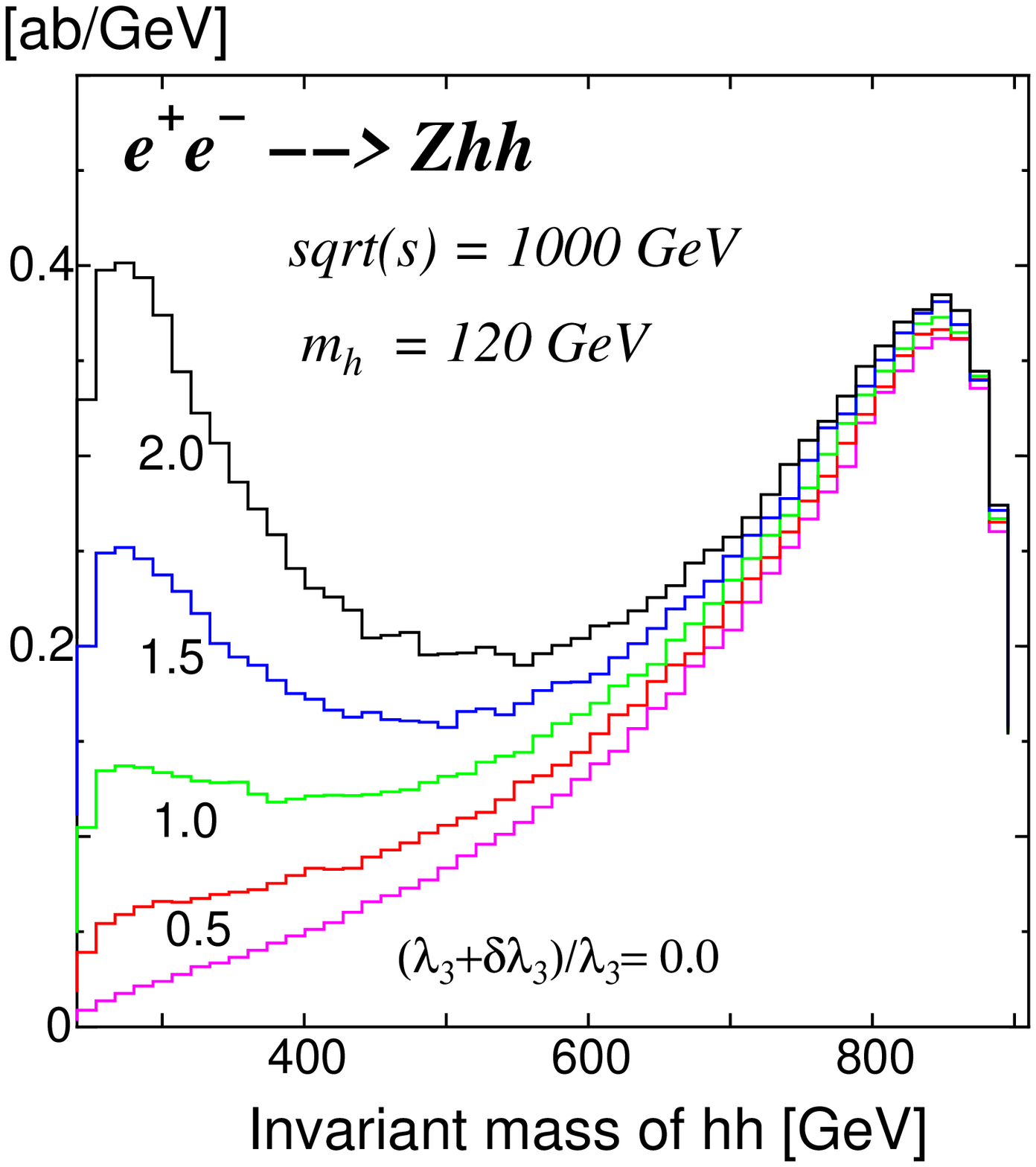} 

Figure 2: The $hh$ imvariant mass dependence 
of the $hhZ$ mode for several values of $\delta\lambda_3$.
\end{minipage}
~~
\begin{minipage}{.6\linewidth}
~~~~~
\epsfxsize=5cm
\epsffile{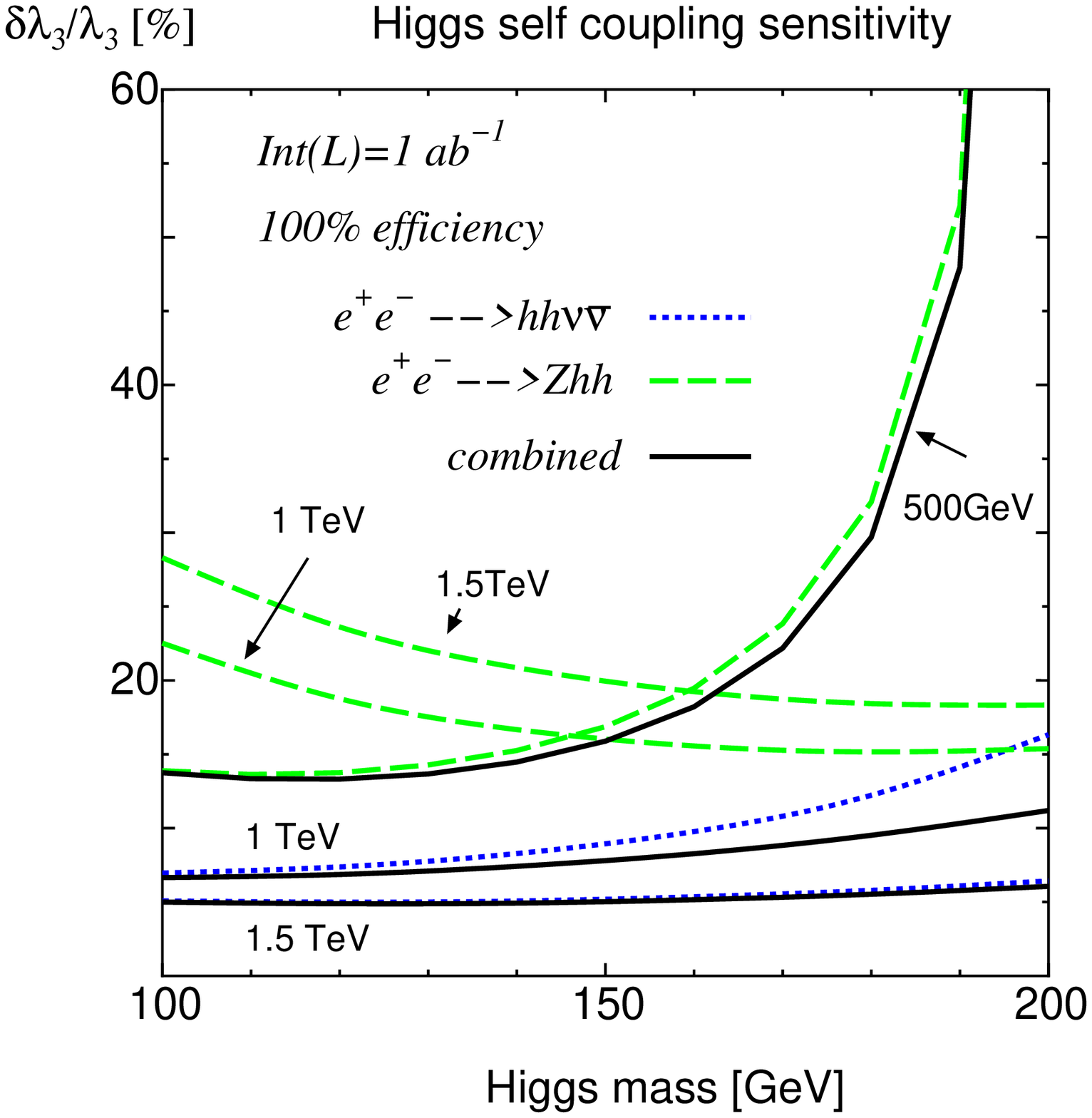} 

Figure 3: The $\lambda_3$ measurement sensitivity; 
$hhZ$ (dashed line),
$hh\nu\bar{\nu}$ (dotted line) and  
combined results (solid line).
\end{minipage}


\vspace{-0.5cm}
\begin{thebibliography}{99}

\vspace{-0.5cm}
\bibitem{BOU}
F.Boudjema and E. Chopin, {\bf Z.Phys. C73} (1996) 85;
V.A.Ilyin, T.Kaneko, Y.Kurihara, A.E.Pukhov and Y.Shimizu,
{\bf Phys.Rev. D54 }(1996) 6717;
Marco Battaglia and Klaus Desch, "Studying the Higgs 
Potential at the e+e- Linear Collider" 
Proceedings of the Linear Collider Workshop LCWS2000; arXiv:hep-ph/0111276.
\bibitem{DJO}A. Djouadi, W. Kilian, M. Muhlleitnera and P.M. Zerwas, 
{\bf Eur.Phys.J. C10} (1999) 45.
\bibitem{KANE} See for example, the SUSY  model was discussed in \cite{DJO}, 
the 2HDM was discussed in S. Kanemura, S. Kiyoura, Y. Okada, 
E. Senaha and C.-P. Yuan, 
Proceedings of the Linear Collider Workshop LCWS2002; arXiv:hep-ph/0209326.
\bibitem{JLC}ACFA Linear Collider Working Group report, 
KEK Report 2001-11; arXiv:hep-ph/0109166.
\bibitem{GRC}J. Fujimoto, T. Ishikawa, M. Jimbo, T. Kaneko, K. Kato, 
S. Kawabata, K. Kon, 
M. Kuroda, Y.Kurihara, Y. Shimizu and H. Tanaka, 
KEK-CP-129; arXiv:hep-ph/0208036.
\bibitem{COMH}A.Pukhov, E.Boos, M.Dubinin, V.Edneral, V.Ilyin,
	D.Kovalenko, A.Kryukov, V.Savrin, S.Shichanin, and A.Semenov,
	INP MSU 98-41/542;
arXiv:hep-ph/9908288

\end{thebibliography}
\end{document}